\documentclass[prb,notitlepage]{revtex4-1}

\usepackage{graphics,graphicx,amsmath}

\newcommand{\ket}[1]{\mbox{$| #1 \rangle$}}

\begin{document}

\title{Double well potentials and quantum gates}

\author{C.\ J.\ Foot}
\affiliation{Department of Physics, University of Oxford, Oxford OX1 3PU, UK}

\author{M.\ D.\ Shotter}
\altaffiliation[Current address: ]{National Institute of Standards and Technology, Gaithersburg, MD 20899, USA}
\affiliation{Department of Physics, University of Oxford, Oxford OX1 3PU, UK}

\begin{abstract}
A system of particles in a double well potential is a
widely studied and useful example for understanding
quantum mechanics. This simple system has recently been used in
theoretical proposals and related experiments as a way to make quantum logic gates for
ultracold atoms confined in optical lattices. Such quantum gates are the fundamental building blocks
for quantum information processing; in these proposals a regular
array of cold atoms in an optical lattice serves as the quantum
register. We explain how this
research can be understood in terms of well-known principles for
systems of identical particles.
\end{abstract}

\maketitle

\section{Introduction}

A system with two particles in a double well potential has been studied in
recent experiments\cite{Anderlini, Trotzky} with ultracold atoms. The atoms were confined in optical lattice potentials, which
are periodic potentials created by standing waves of laser light. These experiments are remarkable because many copies of the
double well system can be prepared simultaneously, each with a well-defined and almost identical quantum state, which is then manipulated. Recently, there has been a
related theoretical proposal to make a quantum logic gate\cite{Hayes} with
neutral atoms held in double well potentials for the purpose of quantum
information processing.

In this paper we explain how the basic
principles underlying this current research can be understood using simple quantum
mechanics. To aid accessibility, we explicitly express the states of the system in
terms of multi-particle wave functions, rather than using the more
compact formulation of creation and annihilation operations as in most of the research papers. To
establish the notation and background, we first consider a system
of two particles in a single well. The reader who is familiar with
this background material can skip to Sec.~\ref{trapLL}.

\section{Two particles in a potential well}

The well-known example of two particles in a potential well
illustrates the important point that the energy levels
of quantum systems do not depend on whether or not the
particles are identical. The occupation of these energy levels is influenced by quantum statistics if, and only if, the
particles are identical.

The Schr\"{o}dinger equation for a system of two particles of mass
$m$ in one dimension is
\begin{equation} \label{SE}
\left[-\frac{\hbar^{2}}{2m}\frac{d^2}{dx^2_1} + V(x_1) -\frac{\hbar^{2}}{2m}\frac{d^2}{dx^2_2}+ V(x_2)\right] \Psi(x_1, x_2)
=E\Psi(x_1, x_2)
\end{equation}
where the potential $V(x)$ is the same for both
particles. If the particles do not interact,
$\Psi$ can be written as the product of the wave functions for the
individual particles
\begin{equation} \label{Psi}
\Psi(x_1, x_2) = \psi(x_1)\psi(x_2) ,
\end{equation}
or linear combinations of such functions.
The lowest configuration of the two-particle system has energy $ E_{g,g}
= 2 E_g$, and wave function $\Psi_{g,g}=\psi_g(x_1)\psi_g(x_2)$, where $\psi_g(x_i)$ is the wave function of the $i$th particle in the ground state. The first excited level of the two-particle system has energy $E_{g,e} = E_g + E_e$, and has two
eigenfunctions (which are a product of single particle states):
\begin{equation} \label{PsiB}
\Psi_{g,e}=\psi_g(x_1)\psi_e(x_2) \mbox{ and } \Psi_{e,g}=\psi_e(x_1)\psi_g(x_2).
\end{equation}
The degeneracy of the first excited state (and higher states)
arises from the symmetry of the system under the exchange of particle labels, that is, the energy of two particles of
the same mass in the same potential well does not depend on which
particle is in the excited state. To determine the effect of an
interaction we need to use degenerate perturbation theory.

\subsection{The effect of an interaction between the particles}

We now consider the effect of an interaction between the particles
so that the Hamiltonian becomes
\begin{equation} \label{Ham}
\hat{H}= -\frac{\hbar^{2}}{2m}\frac{d^2}{dx^2_1}
-\frac{\hbar^{2}}{2m}\frac{d^2}{dx^2_2}+ V(x_2)+ V(x_1) + V_{1,2}(x_1 - x_2) .
\end{equation}
The potential $V_{1,2} (x_1 -x_2)$ represents the interaction
between the two particles, which we consider as a perturbation.
We only consider potentials that are a function of the distance
between the two particles $|x_1-x_2|$, for example, the Coulomb potential
between charged particles.

At this stage those who have encountered similar situations,
for example, in solving the helium atom, are inclined to anticipate the
result and exploit the freedom arising from the degeneracy to
rewrite the two-particle wave functions as linear combinations of the
original wave functions that are symmetric or antisymmetric under
exchange of particle labels. This approach may be prompted by the knowledge that both spatial and spin wavefunctions are either symmetric or antisymmetric for systems of identical particles.
However, for now we consider only
distinguishable particles to see the natural emergence of symmetric
and antisymmetric states, without explicitly invoking the exchange symmetry operator.

The effect of $V_{1,2}(x_1 -x_2)$ on the non-degenerate ground
state is given by the first-order perturbation:
\begin{equation}\label{Egg}
\Delta E_{g,g} = \!\int\! dx_1\!\int\! dx_2 \left|\psi_g(x_1) \right|^2 \left|\psi_g(x_2)\right|^2 V_{1,2}(x_1 -x_2).
\end{equation}
To evaluate the effect of the perturbation on the degenerate states
of Eq.~(\ref{PsiB}) we write the general wave function in the basis of
the two states as
\begin{equation}\label{Psivec}
\Psi(x_1, x_2) \simeq a \psi_g(x_1) \psi_e(x_2) + b \psi_e(x_1) \psi_g(x_2).
\end{equation}
This projection of the wave function onto two states is a good
approximation when the perturbation is sufficiently small that it
does not mix in significant amplitudes of other states. It is
convenient to represent these two degenerate basis states as
vectors:
\begin{equation}\label{basisvec}
\psi_g(x_1) \psi_e(x_2) \mapsto \begin{pmatrix} 1 \\
0 \\
\end{pmatrix} \mbox{ and } \psi_e(x_1) \psi_g(x_2) \mapsto \begin{pmatrix}
0 \\
1 \\
\end{pmatrix}.
\end{equation}
In this truncated basis the Schr\"{o}dinger equation $\hat{H} \Psi
=E\Psi $ reduces to a matrix equation (with eigenvalue $E$):
\begin{equation}\label{HamMx}
\begin{pmatrix}
E_{g,e}+J & K \\
K & E_{g,e}+J \\
\end{pmatrix}
\begin{pmatrix}
a \\
b \\
\end{pmatrix} = E
\begin{pmatrix}
a \\
b \\
\end{pmatrix}.
\end{equation}
The eigenvalues are determined by the roots of the equation
\begin{equation}\label{EigenV}
(E_{g,e}+J-E)^2-K^2=0,
\end{equation}
which are
\begin{equation}\label{EigenE}
E = E_{g,e} +J \pm K.
\end{equation}
Here $J$ is the `direct' integral given by
\begin{equation}\label{direct}
J = \!\int\! dx_1\! \int\! dx_2~ \left| {\psi_{g}(x_1)} \right|^2 \big| {\psi_{e}(x_2)} \big|^2 V_{1,2}(x_1-x_2),
\end{equation}
and $K$ is the `exchange' integral. (It might seem strange to
call $K$ the exchange integral when the particles are
distinguishable, but it has the same form for identical
particles.) The exchange integral is given by
\begin{equation}\label{exchange}
K = \!\int\! dx_1\!\int\! dx_2\,\psi_g(x_1)\psi^*_e(x_1)\psi^*_g(x_2)\psi_e(x_2) V_{1,2}(x_1 -x_2).
\end{equation}
$K$ is real (the symmetry of the integral for interchange
of variables $x_1\leftrightarrow x_2$ implies that $K^*=K$). Unlike
the direct integral, the exchange integral $K$ cannot be interpreted
in terms of a simple classical picture; attempts to explain the exchange integral
classically lead to nonsensical concepts such as `exchange forces'. (For example, the
gross structure of the helium atom is determined by electrostatic
interactions and nothing else.)

The two eigenvectors and corresponding spatial wave functions are
\begin{align}
\begin{pmatrix}
a \\
b \\
\end{pmatrix} =
\begin{pmatrix}
1/\sqrt{2} \\
-1/\sqrt{2} \\
\end{pmatrix} & \mapsto \dfrac{1}{\sqrt 2} \left[ \psi_g(x_1) \psi_e(x_2) - \psi_e(x_1) \psi_g(x_2) \right] \
&& \mathrm{for}~E=E_{g,e}+J-K
\label{Minus}\\
\begin{pmatrix}
1/\sqrt{2} \\
1/\sqrt{2} \\
\end{pmatrix} & \mapsto \dfrac{1}{\sqrt 2} \left[ \psi_g(x_1) \psi_e(x_2) + \psi_e(x_1)
\psi_g(x_2) \label{Plus}
\right] && \mathrm{for}~E=E_{g,e} +J+K.
\end{align}
We see that the eigenstates of the Hamiltonian with an interaction between the
distinguishable particles are the antisymmetric and symmetric
combinations of the original product wave functions. Note that the form of these eigenstates has been derived without reference
to the intrinsic exchange symmetry of indistinguishable particles. Instead, this behavior may be understood as a direct consequence of the symmetry of the Hamiltonian with respect to the exchange of the particle labels $x_{1}$ and $x_{2}$.

For the corresponding problem with indistinguishable particles, that is, particles that possess an intrinsic exchange symmetry, the only difference is in the occupation of the two states of Eqs.~(\ref{Minus}) and (\ref{Plus}). Particles with a certain exchange symmetry and a spin wavefunction of a certain exchange parity can occupy only one of the two states of Eqs.~(\ref{Minus}) and (\ref{Plus}); the other is forbidden. Nevertheless, the wave functions and energies of the two states remain the same as for distinguishable particles.

\subsection{A delta-function interaction}

The interaction between two ultracold neutral atoms can be well
approximated by a delta-function potential
\begin{equation}\label{Vdelta}
V_{1,2}(x_1- x_2) = a \delta(x_2 - x_1)
\end{equation}
where $a$ is a constant, and $a>0$ for a repulsive interaction. For
this so-called contact interaction, the direct and exchange
integrals are the same:
\begin{equation}\label{directdelta}
J = K = \!\int\! dx~ \left|\psi_g(x) \right|^2 \big|\psi_e(x)\big|^2 a
\end{equation}
Because $J-K=0$, the interaction does not change the energy of the
antisymmetric wave function in Eq.~(\ref{Minus}), which is an
consequence of the function being equal to zero when $x_1 =
x_2$. A similar argument is often used
to explain why the triplet terms of the helium atom are lower in energy
than the singlets, namely that the expectation
value of the electrostatic repulsion between the two electrons is
lower for the spatial wave function in which the two parts have
opposite sign.

\section{Identical particles and spin statistics}

If the two particles are identical, their spin has an influence on
the occupancy of the energy levels. In particular, the requirements
of overall exchange antisymmetry for fermions and symmetry for
bosons introduces a connection between the spin and spatial wave
functions. Consequently states with different spin have different
energies, for example, the singlet and triplet terms in helium. It is
tempting to think that the energies themselves depend on spin,
but they do not. The wave functions for two fermions are linear combinations of the terms
\begin{equation}\label{Psifermions}
\Psi_{\mathrm{space}}^{\mathrm{(S)}}\Psi_{\mathrm{spin}}^{\mathrm{(A)}} \mbox{ and } \Psi_{\mathrm{space}}^{\mathrm{(A)}}\Psi_{\mathrm{spin}}^{\mathrm{(S)}},
\end{equation}
where (A) denotes antisymmetric with respect to exchange of the
particle labels $1\leftrightarrow 2$, and (S) denotes symmetric.
This subset of all the possible two-particle wave functions has
overall antisymmetry with respect to the exchange of particle labels as
required for identical fermions. The spin wave functions are
given in Table~\ref{Table1}. There is a singlet for
$\Psi_{\mathrm{spin}}^{\mathrm{(A)}}$ and a triplet for
$\Psi_{\mathrm{spin}}^{\mathrm{(S)}}$ making a total of four functions.

A well-known example of these spin states occurs in the helium atom: for the
ground configuration (1s$^2$) there is only a symmetric spatial wave
function $\Psi_{\mathrm{space}}^{\mathrm{(S)}}=
\psi_{\mathrm{1s}}^2$ associated with a spin singlet. The first
excited configuration (1s2s) is split into two terms: a singlet
($^1\mathrm{S}$) and a triplet ($^3\mathrm{S}$) whose energy
separation is twice the exchange integral given by
\begin{equation} \label{e3exchange}
K=\frac{e^{2}}{{4\pi \epsilon _{0}}}\!\int\! d{\mathbf r}_1^{3}\!\int\!
d{\mathbf r}_2^{3}\ \psi_{\mathrm{1s}}^{*}(r_1)\psi_{\mathrm{2s}}^{*}(r_2) \frac{{1}}{| \mathbf r_2- \mathbf r_1 |} \psi_{\mathrm{2s}}(r_1)
\psi_{\mathrm{1s}}(r_2),
\end{equation}
where $e^2/ 4\pi \epsilon_0 {| \mathbf r_2- \mathbf r_1 |}$ is the
Coulomb repulsion between the electron at $\mathbf r_1$ and the
electron at $\mathbf r_2$.\cite{Foot}
For two bosons the wave function is symmetric with respect to
exchange of the particle labels, so that the wave functions are linear combinations of the terms
\begin{equation}\label{Psibosons}
\Psi_{\mathrm{space}}^{\mathrm{(S)}}\Psi_{\mathrm{spin}}^{\mathrm{(S)}} \mbox{ and } \Psi_{\mathrm{space}}^{\mathrm{(A)}}\Psi_{\mathrm{spin}}^{\mathrm{(A)}}.
\end{equation}
This brief review of the physics of two particles in a single well
serves as an introduction for the treatment of a double well system in
the following.

\section{Trapping atoms using light} \label{trapLL}

Many experiments in ultracold atomic physics use laser light to trap atoms. The laser light interacts with an atomic resonance to form a spatially varying potential, which can confine the atoms. Consider a laser beam incident on a two-level atom, which has unperturbed eigenenergies of $E_{g}$ in the ground state and $E_{e}$ in the excited state. These energies refer to electronic (internal) states of the atoms rather than the motional states of the atomic center of mass. The laser beam has an oscillating electric field ${\mathcal{E}}(t)={\mathcal{E}}_{0}\cos(\omega_{l}t)$, which introduces a time-dependent coupling between the ground and excited states of the atom. The coupled Schr\"{o}dinger equation for the ground and excited electronic states of the atom in the presence of the oscillating electric field of the laser beam can be shown to be\cite{Foot}
\begin{subequations}
\label{thiseq}
\begin{align}
i\hbar\frac{\partial \psi_{g}}{\partial t}&=E_{g}\psi_{g}+\hbar \Omega \cos(\omega_{l}t) \psi_{e}\\
i\hbar\frac{\partial \psi_{e}}{\partial t}&=E_{e}\psi_{e}+\hbar \Omega \cos(\omega_{l}t) \psi_{g},
\end{align}
\end{subequations}
where $\Omega$ is a time-independent parameter proportional to ${\mathcal{E}}_{0}$.

To solve for the steady state of Eq.~\eqref{thiseq} we make a time-dependent substitution for the wavefunction
\begin{subequations}
\begin{align}
\psi_{g}&=\tilde{\psi}_{g}\exp[-iE_{g}t/\hbar]\\
\psi_{e}&=\tilde{\psi}_{e}\exp[-i(E_{g}+\hbar \omega_{l})t/\hbar],
\end{align}
\end{subequations}
so that
\begin{subequations}
\label{RWAall}
\begin{align} \label{RWA}
i\frac{\partial \tilde{\psi}_{g}}{\partial t}&= \frac{\Omega}{2} (1+e^{-2i\omega_{l}t}) \tilde{\psi}_{e}\\
i\frac{\partial \tilde{\psi}_{e}}{\partial t}&=-\delta \tilde{\psi}_{e}+ \frac{\Omega}{2} (1+e^{2i\omega_{l}t}) \tilde{\psi}_{g}, \label{RWA2}
\end{align}
\end{subequations}
where $\hbar \delta=\hbar \omega_{l}-(E_{e}-E_{g})$.

When the frequency detuning of the laser light from the transition frequency $\delta$ and the laser coupling parameter $\Omega$ are much less than the transition frequency $(E_{e}-E_{g})/\hbar$, Eq.~(\ref{RWAall}) can be simplified. The rapidly oscillating terms $e^{\pm 2i\omega_{l}t}$ vary with a much shorter period than all other time scales in the problem. Therefore the integral of a term with a rapidly varying phase factor such as $e^{2i\omega_{l}t} \tilde{\psi}_{g}$ is much less than the integral of the term without this factor $\tilde{\psi}_{g}$. Hence we can discard terms from the coupled equations with a rapidly varying phase. This is called the rotating wave approximation; it turns out to be a very good approximation for laser-atom interactions. The resulting effective time-independent Hamiltonian in matrix form is
\begin{equation}
H_{\rm eff}=\hbar \begin{pmatrix}
0 & \Omega/2 \\
\Omega/2 & -\delta \\
\end{pmatrix}.
\end{equation}

Diagonalizing this Hamiltonian gives the energy shift of the atomic states due to the laser beam. The frequency detuning $\delta$ of the laser light used in experiments is typically much greater than the coupling parameter $\Omega$ in order to minimize the spontaneous scattering of the laser light from the trapped atoms. In this limit the energy shifts for the ground and excited state energy levels due to the laser light are
\begin{equation}
\Delta E_{g}=\frac{\hbar\Omega^{2}}{4\delta} \mbox{ and } \Delta E_{e}=-\frac{\hbar\Omega^{2}}{4\delta}.
\end{equation}

This energy shift is often called a dipole energy shift or an AC Stark shift. As $\Omega$ is proportional to the electric field ${\mathcal{E}}_{0}$, the energy shifts are proportional to the intensity of the laser beam. The intensity of laser beams can vary spatially so that the dipole energy shift can be used to trap atoms within a potential. For example, when the frequency of the laser beam is less than the frequency of the atomic transition ($\delta < 0$), ground state atoms are trapped at laser intensity maxima, such as the focus of a single laser beam.

The simple relation between the atomic potential and laser intensity $V(\mathbf{x})\propto I(\mathbf{x})$ is the basis for the trapping of ultracold atoms using lasers. All manner of light patterns can be created to form potential landscapes
for the ultracold atoms. For example, a standing wave arising from the
interference of two counter-propagating laser beams creates a periodic
potential of hills and valleys for the atoms called an optical lattice. These periodic
potentials can be extended to two and three dimensions using additional
pairs of laser beams. A double well trap, such as discussed in this paper, may be formed using two independently focused laser beams. The laser beams, and hence the potential that the atoms experience, can be altered in real-time, making laser dipole traps a highly flexible experimental tool with which to manipulate ultracold atoms.

\section{Double well potential}\label{sectDouble}

\subsection{A single particle in a double well}

\begin{figure}[h!]
\centering
\includegraphics[scale=0.7]{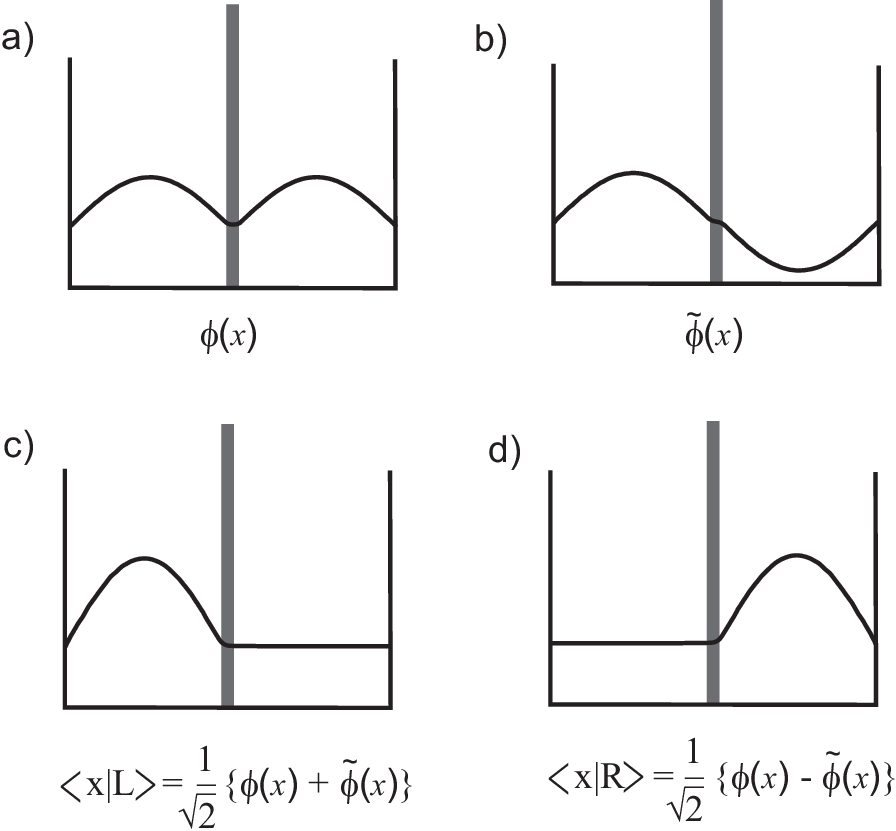}
\caption{The wave functions of a single particle in a double well
potential are illustrated using a square well with a high barrier
for simplicity. (a) The wave function $\phi(x)$ of lowest energy has
even parity. (b) The energy of the lowest wave function with odd parity
$\tilde\phi(x)$ is slightly higher than that of $\phi(x)$. The
energy difference between these two states increases as the tunneling rate
increases. (c) and (d) show the combinations of $\phi$ and
$\tilde\phi$ given in Eq.~(\ref{xLR}) and represent the
particle localized in the left and right well respectively.} \label{Figure1}
\end{figure}

We consider double wells that have symmetry with respect to
inversion ($x\rightarrow -x$), so that the eigenfunctions of
particles in this potential have definite parity. (The
Hamiltonian of the system commutes with the parity operator
$\hat{P}$, that is, $[\hat{H}, \hat{P}]=0$, and hence these operators have
simultaneous eigenfunctions.) The even and odd parity wave functions
for a single particle in a double square well potential with a thin
barrier are illustrated in Fig.~\ref{Figure1}.

The energy of the state whose wave function $\phi$ has even parity is less
than that of the wave function with odd parity $\tilde\phi$; the
barrier has less effect on the odd wave function because $\tilde\phi$
has a node at $x=0$ no matter the height of the barrier,
whereas the energy of the even parity wave function decreases as the
barrier height decreases. For the simple case of square
well potentials, when the barrier is reduced from $\infty$ to zero,
so that the well becomes twice its original length, the ground state
energy reduces to a quarter of its original value. A
qualitatively similar argument holds for any double well with
reflection symmetry: the solution of the Schr\"{o}dinger
equation with no nodes has the lowest eigenenergy.

We can form linear superpositions of the energy eigenstates
\begin{subequations}
\label{xLR}
\begin{align}
\langle x | L\rangle &= \dfrac{1}{\sqrt 2}\left[ \phi(x)+ \tilde\phi(x) \right] \label{xL}\\
\langle x | R\rangle &= \dfrac{1}{\sqrt 2}\left[ \phi(x)- \tilde\phi(x) \right], \label{xR}
\end{align}
\end{subequations}
which correspond to wave functions in which the particle is located
predominantly in the left or right well respectively, as
illustrated in Fig.~\ref{Figure1}. These are not eigenstates of the system, and thus
the wave function of a particle initially prepared in the left well
at $t=0$ evolves according to
\begin{subequations}
\begin{align}
\psi(x,t) &= \dfrac{1}{\sqrt 2}\left[ \phi(x) e^{- i E t/\hbar} + \tilde\phi(x)e^{- i \tilde E t/\hbar} \right] \\
&= \dfrac{1}{\sqrt 2} e^{\mathrm -i E t/\hbar} \left[ \phi(x) + \tilde\phi(x)e^{- i \Delta E t/\hbar} \label{iJt}
\right],
\end{align}
\end{subequations}
where $\Delta E /\hbar = (\tilde E - E)/\hbar$ is the tunneling rate between
the wells. If $\Delta E t /\hbar = 2 \pi n$ for integer $n$, the wave
function is proportional to $ \langle x | L\rangle $, and if
$\Delta E t /\hbar = (2n-1) \pi$, the wave function is proportional to $
\langle x | R\rangle $, corresponding to the particle having tunneled
into the other well. For negligible tunneling between the
wells (very high barrier) it is possible to describe the particle as
being either in the left or right well. This is clearly the case when
the wells are very far apart so that the particle is localized in
one of the wells, and is the situation at the start of the
experiments we shall describe.

\subsection{Two interacting particles in a double well}

The previous sections have prepared all the tools we
need to understand the behavior of two particles in a double well.
We choose a potential with a high barrier so that initially the
system is close to the limit of two separate wells. There is an
antisymmetric spatial wave function
$\Psi_{\mathrm{space}}^{\mathrm{(A)}}$ which can be expressed in
terms of the left/right wave functions of the individual particles
as
\begin{equation}
\Psi_{\mathrm{space}}^{\mathrm{(-)}} (x_1, x_2) = \dfrac{1}{\sqrt 2} \left[ \langle x_1 | L\rangle \langle x_2 | R\rangle - \langle x_1 | R\rangle \langle x_2 | L\rangle \right].
\label{this2}
\end{equation}
Equation~\eqref{this2} can be rewritten using the ket notation to represent $\Psi_{\mathrm{space}}^{\mathrm{(-)}}$
\begin{equation}\label{PsiLRRL}
\Psi_{\mathrm{space}}^{\mathrm{(-)}} = \dfrac{1}{\sqrt 2} \left[|LR \rangle- |RL \rangle
\right].
\end{equation}
Similarly, the symmetric wave functions
($\Psi_{\mathrm{space}}^{\mathrm{(S)}}$) are linear combinations of the three terms
\begin{equation}
|LL \rangle,\ |RR \rangle, \mbox{ and }
\Psi_{\mathrm{space}}^{\mathrm{(+)}} = \dfrac{1}{\sqrt 2} \left[
|LR \rangle+ |RL \rangle \right].
\end{equation}
Writing the four linearly independent spatial wave functions in this way highlights their
similarity with the form of the four spin wave functions for two
spin one-half particles; both have one $\Psi^{\mathrm{(A)}}$ and three
$\Psi^{\mathrm{(S)}}$. We combine space and spin functions with the
restriction of overall antisymmetry, as in Eq.~(\ref{Psifermions}), and
find by inspection of Table~\ref{Table1} that there are six
two-particle wave functions for identical spin one-half particles.

\begin{figure}[h!]
\includegraphics[scale=0.8]{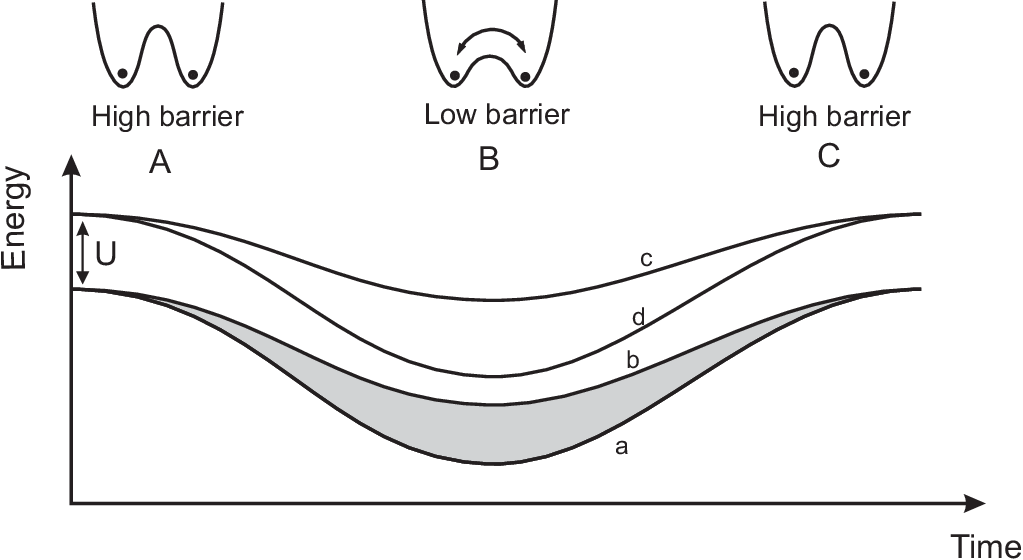}
\caption{Schematic diagram of the operation of a quantum gate for
two atoms in a double well potential. The energies of the four
lowest states (see Table~\ref{abcdstates}) of the two-atom double well system are
shown as a function of time as the height of the barrier is lowered
and then raised. For the initial and final conditions (marked A and
C) tunneling is negligible, whereas during operation (for example, at B)
the barrier is much lower; the barrier could even be reduced to
zero so that both atoms are in a single well for a certain time
before the barrier is raised. The phase $\Delta \varphi$ of the gate
is the integral of the shaded region between paths a and b (divided
by $\hbar$). The barrier is altered slowly and smoothly to ensure
the adiabatic evolution of the states along the given paths
(otherwise states a and c mix). The condition for adiabatic
evolution is that the gate must take place over a time
significantly longer than $h/U$. The states have been labeled to be
consistent with Ref.~\onlinecite{Hayes}.} \label{adiabatic_evolution}
\end{figure}

Now consider the effect of a repulsive contact interaction between
the particles. This interaction has little effect on the energy
when the two particles are localized in different wells (with small
overlap of their wave functions), that is, for
$\Psi_{\mathrm{space}}^{(-)}$ and $\Psi_{\mathrm{space}}^{(+)}$. However, for the wave functions containing
$\psi_R(x_1)\psi_R(x_2) $ ($|RR \rangle$) and
$\psi_L(x_1)\psi_L(x_2) $ ($|LL \rangle$), the interaction raises
the energy by an amount which we denote as $U$. The
integral for $U$ resembles Eq.~(\ref{directdelta}):
\begin{equation}\label{U}
U = \!\int\! dx~ \left|\psi_R(x) \right|^4 a,
\end{equation}
and analogously for $\psi_L(x)$. Figure~\ref{adiabatic_evolution} and Table~\ref{abcdstates} show the eigenenergies
and associated fermionic eigenstates for a double well with a repulsive interaction.

In the following we shall mainly consider the subset of states with
the particles in different wells. These are separated by the energy
gap $U$ from the higher lying states $|L L \rangle \chi^{(-)}$
and $|R R \rangle \chi^{(-)}$.

\section{Quantum gate for two fermions}

In quantum computing notation the state in which both particles are
spin-up $|\uparrow\uparrow\rangle$ can be written as $|00 \rangle$
and conversely $|\downarrow\downarrow\rangle\equiv |11 \rangle$. The
state of the system is encoded in terms of the four basis states as
\begin{equation}
\Psi = a |00 \rangle + b |01 \rangle + c |10 \rangle + d |11 \rangle,
\end{equation}
where $a$, $b$, $c$, and $d$ are complex coefficients. The state $|00
\rangle$ corresponds to a spin-up particle in each well for which
the antisymmetrized wave function (unnormalized) is
\begin{equation}\label{PsiOO}
|00 \rangle \propto \langle x_1 | L\rangle |\uparrow_1 \rangle
\langle x_2 | R\rangle |\uparrow_2\rangle - \langle x_1 |
R\rangle|\uparrow_1\rangle \langle x_2 | L\rangle
|\uparrow_2 \rangle.
\end{equation}
Equation~\eqref{PsiOO} can be expressed in terms of the energy eigenstates
as
\begin{subequations}
\begin{align}
|00\rangle &\propto \left(| LR \rangle - |RL \rangle \right)
|\uparrow\uparrow\rangle,\\
~ &\propto \Psi_{\mathrm{space}}^{(-)}|\uparrow\uparrow\rangle .
\label{state00}
\end{align}
\end{subequations}
The state $|01 \rangle$ corresponds to spin-up in the left well and
spin-down on the right for which the antisymmetrized (unnormalized) wave function
is
\begin{equation}\label{Psi01}
| 01 \rangle \propto \langle x_1 | L\rangle |\uparrow_1 \rangle
\langle x_2 | R\rangle |\downarrow_2 \rangle - \langle x_1 |
R\rangle|\downarrow_1\rangle \langle x_2 | L\rangle
|\uparrow_2\rangle.
\end{equation}
Equation~\eqref{Psi01} can be expressed in terms of the energy eigenstates
as
\begin{subequations}
\begin{align}
|01\rangle &\propto | LR \rangle |\uparrow\downarrow\rangle - |RL
\rangle |\downarrow\uparrow\rangle \\
&\propto \left(\Psi_{\mathrm{space}}^{(-)} +
\Psi_{\mathrm{space}}^{(+)} \right) |\uparrow\downarrow \rangle +
\left(\Psi_{\mathrm{space}}^{(-)} - \Psi_{\mathrm{space}}^{(+)} \right) |\downarrow\uparrow\rangle\\
&\propto \Psi_{\mathrm{space}}^{(-)} \left(
|\uparrow\downarrow \rangle + |\downarrow\uparrow \rangle \right)
+ \Psi_{\mathrm{space}}^{(+)} \left(|\uparrow\downarrow \rangle
- |\downarrow\uparrow \rangle \right),
\end{align}
\end{subequations}
and similarly for $ |10 \rangle$ and $ |11 \rangle$. Thus we write
the quantum computational basis states in terms of the four eigenstates
of two spin one-half particles (which are degenerate for zero tunneling) as
\begin{subequations}
\label{basis}
\begin{align}
|00\rangle &= \Psi_{\mathrm{space}}^{(-)}|\uparrow\uparrow\rangle\\
|01\rangle &= \left(\Psi_{\mathrm{space}}^{(-)}\chi^{(+)} + \Psi_{\mathrm{space}}^{(+)}\chi^{(-)} \right) /\sqrt 2\\
|10\rangle &= \left( \Psi_{\mathrm{space}}^{(-)}\chi^{(+)} - \Psi_{\mathrm{space}}^{(+)}\chi^{(-)} \right) /\sqrt 2 \\
|11\rangle &=
\Psi_{\mathrm{space}}^{(-)}|\downarrow\downarrow\rangle .
\end{align}
\end{subequations}

The barrier between the wells can be altered by changing the intensity or separation of the two laser beams used to form the double well potential. When the barrier is lowered slowly, $\Psi_{\mathrm{space}}^{(-)}$ and
$\Psi_{\mathrm{space}}^{(+)}$ adiabatically evolve into
$\Psi_{\mathrm{a}}$ and $\Psi_{\mathrm{b}}$ respectively, which are
no longer degenerate. Further details are given in Ref.~\onlinecite{Hayes} (see also Table~\ref{abcdstates}). To understand the operation of
the gate all we need to know is that a phase difference accumulates
between these states because they have different eigenenergies during the
gate operation. Thus if the barrier is lowered for a certain interval
and then raised again (to switch off the tunneling), the wave
function that starts as $|01\rangle$ becomes
\begin{equation}\label{evolves}
\Psi(t) = e^{ -i\varphi} \left(
\Psi_{\mathrm{space}}^{(-)}\chi^{(+)} + e^{ i\Delta\varphi}
\Psi_{\mathrm{space}}^{(+)}\chi^{(-)} \right) /\sqrt 2.
\end{equation}
Controlling the process so that the accrued phase shift $\Delta
\varphi = \pi$ causes this state to evolve into
\begin{subequations}
\begin{align}
\Psi(t_\pi) &= e^{ -i\varphi}\left( \Psi_{\mathrm{space}}^{(-)}\chi^{(+)} + e^{i \pi } \Psi_{\mathrm{space}}^{(+)}\chi^{(-)} \right) /\sqrt 2\\
&= e^{ -i\varphi} \left( \Psi_{\mathrm{space}}^{(-)}\chi^{(+)} - \Psi_{\mathrm{space}}^{(+)}\chi^{(-)} \right) /\sqrt 2
\\
&= e^{ -i\varphi} |10 \rangle .
\end{align}
\end{subequations}
The global phase factor $e^{ -i\varphi}$ is unimportant. All other combinations also pick up this phase, and therefore it does not affect the relative phase.

It can be seen that a system initialized in
$|01\rangle$ evolves into $|10 \rangle$ and \textit{vice versa},
which implements the SWAP operation $|01 \rangle \leftrightarrow
|10 \rangle$. When $\Delta\varphi = 2\pi$, the system cycles back to
its initial state. More generally, the states $|01\rangle$ and $
|10 \rangle$ evolve into a coherent superposition, while the $|00\rangle$ and $
|11 \rangle$ states remain unaffected. The SWAP
operation can be written as the $4\times 4$ matrix
\begin{equation}
U_{\rm SWAP} = \begin{pmatrix}
1 & 0 & 0 & 0 \\
0 & 0 & 1 & 0 \\
0 & 1 & 0 & 0 \\
0 & 0 & 0 & 1 \\
\end{pmatrix}
\end{equation}
for the basis states in Eq.~(\ref{basis}). The SWAP gate cannot be used to entangle qubits. The operation on two unentangled qubits is
\begin{equation}
(\alpha |0 \rangle + \beta |1 \rangle)(\gamma |0 \rangle + \delta |1 \rangle)\equiv \begin{pmatrix}
\alpha \gamma \\
\alpha \delta \\
\beta \gamma \\
\beta \delta \\
\end{pmatrix}
\xrightarrow{\textrm{SWAP}}
\begin{pmatrix}
\alpha \gamma \\
\beta \gamma \\
\alpha \delta \\
\beta \delta \\
\end{pmatrix}\equiv(\gamma |0 \rangle + \delta |1 \rangle)(\alpha |0 \rangle + \beta |1 \rangle),
\end{equation}
which (as the name implies) swaps the state of the qubits without entangling them.

A closely related gate called the ``square-root-of-SWAP'' gate can perform a useful
entangling operation in quantum computation. This operation has the property
\begin{equation}\label{eqSW}
U_{\sqrt{\mathrm{SWAP}}}U_{\sqrt{\mathrm{SWAP}}}=U_{\mathrm{SWAP}}.
\end{equation}
A matrix satisfying this condition is
\begin{equation}\label{rSW}
U_{\sqrt{\mathrm{SWAP}}} = \begin{pmatrix}
1 & 0 & 0 & 0 \\
0 & \frac12(1+i) & \frac12(1-i) & 0 \\
0 & \frac12(1-i) & \frac12(1+i) & 0 \\
0 & 0 & 0 & 1 \\
\end{pmatrix}.
\end{equation}
To implement the $\sqrt{\mathrm{SWAP}}$ gate of Eq.~(\ref{rSW}) the system undergoes half of the phase evolution of the SWAP gate.
A square-root-of-SWAP gate cannot be broken down into a
combination of single qubit gates, that is, gates operating on one
qubit at a time. Furthermore, it entangles two unentangled qubits giving the resultant state [from Eq.~(\ref{rSW})]
\begin{equation}
(\alpha |0 \rangle + \beta |1 \rangle)(\gamma |0 \rangle + \delta |1 \rangle)\equiv\begin{pmatrix}
\alpha \gamma \\
\alpha \delta \\
\beta \gamma \\
\beta \delta \\
\end{pmatrix}
\xrightarrow{U_{\sqrt{\textrm{SWAP}}}}
\begin{pmatrix}
\alpha \gamma \\
\frac12 (1+i) \alpha \delta+ \frac12(1-i) \beta \gamma \\
\frac12 (1+i) \beta \gamma + \frac12(1-i) \, \alpha \delta \\
\beta \delta \\
\end{pmatrix},
\end{equation}
which cannot be written as a product of two single qubit states.

A square-root-of-SWAP gate,
used alongside single-qubit operations, is
sufficient to provide a universal set of gates for quantum
computation.\cite{Nielsen} Single qubit gates would be implemented in this system by the rotation of the atomic spin
orientation of the atom in one of the wells, which may be accomplished simply
by optical (Raman) or microwave pulses. The implementation of these powerful
ideas with fermionic atoms in optical lattices is described in Ref.~\onlinecite{Hayes}. Similar principles can also be applied
to a system of two bosons (selecting just two out of the possible
spin states to form a suitable basis). More details of a gate for
bosons are given in Refs.~\onlinecite{Vaucher, Anderlini, Trotzky}.

\section{Conclusion}

We have shown how recent experimental and theoretical research on
ultracold atoms in double well potentials can be simply understood. The experiments have been carried out
with bosonic atoms (the isotope $^{87}$Rb) for technical reasons, but for
simplicity the fermionic case has been used to expound the
principles. This system described in Ref.~\onlinecite{Hayes} is particularly intriguing for the way in which the
influence of the nuclear spin on the spatial wave functions is
exploited.

Second quantized notation tends to be
used for a more general solution, as in most of the papers we have cited, because this notation becomes more
efficient in keeping track of exchange symmetry as the number of
particles in the system increases.

There is great current interest
in using cold atoms in optical lattices for quantum information processing, including the direct quantum
simulation of strongly correlated many particle systems (such as
those in condensed matter physics). The simple approach outlined in
this paper gives an
intuitive way of understanding aspects of these systems.\\

The authors would like to thank the referees and editors for helpful comments. M.S. would also like to thank Christ Church (Oxford), NIST (Gaithersburg) and the Lindemann Trust for financial assistance.\\

\clearpage\section*{Tables}

\begin{table}[h!]
\centering
\begin{tabular}{|c|c|}
\hline
$\underline{\Psi_{\mathrm{space}}^{\mathrm{(A)}}}$ & $\underline{\Psi_{\mathrm{spin}}^{\mathrm{(A)}}}$ \\
$ \dfrac{1}{\sqrt 2} \left[|LR \rangle -|RL \rangle \right]\equiv \Psi_{\mathrm{space}}^{(-)}$ &
$ \dfrac{1}{\sqrt 2} \left[|\uparrow\downarrow \rangle - |\downarrow\uparrow\rangle \right] \equiv \chi^{(-)}$ \\
\hline
$\underline{\Psi_{\mathrm{space}}^{\mathrm{(S)}}}$ & $\underline{\Psi_{\mathrm{spin}}^{\mathrm{(S)}}}$ \\
$ |\,LL\,\rangle$ & $|\uparrow\uparrow\,\rangle $ \\
$ \dfrac{1}{\sqrt 2} \left[|LR \rangle +|RL \rangle \right]\equiv \Psi_{\mathrm{space}}^{(+)}$ &
$ \dfrac{1}{\sqrt 2} \left[ |\uparrow\downarrow \rangle + |\downarrow\uparrow\rangle \right] \equiv \chi^{(+)} $ \\
$ |\,RR\,\rangle$ & $|\downarrow\downarrow\,\rangle $ \\
\hline
\end{tabular}
\caption{The spatial and spin wave functions for two
particles can either be symmetric (S) or antisymmetric (A) with respect to
exchange of the particle labels. For a pair of fermions the spatial and spin wave functions can be combined to form six linearly independent two-particle wave functions of the form
$\Psi_{\rm space}^{\rm (A)}\Psi_{\rm spin}^{\rm (S)}$
or
$\Psi_{\rm space}^{\rm (S)}\Psi_{\rm spin}^{\rm (A)}$.
To form these states, each of the six symmetric functions in the bottom row is
associated with one of the functions in the top row. This subset (of
the sixteen possible product states) has overall antisymmetry with
respect to exchange of particle labels as required for identical
fermions.}
\label{Table1}
\end{table}

\begin{table}[h!]
\begin{tabular}
{|c|p{1.9cm}|c|c|p{4.8cm}|p{5cm}|}\hline
State & Spin states & Exchange symmetry & Parity & Eigenfunctions for a high barrier & Eigenfunctions for a low barrier \\
\hline
a & singlet & $+$ & $+$ & $\dfrac{1}{\sqrt 2} \left[ |LR\rangle +|RL \rangle \right] \equiv \Psi_{\mathrm{space}}^{(+)}$ & \
$\ket{\phi \phi}$ \\
b & triplet & $-$ & $-$ & $ \dfrac{1}{\sqrt 2} \left[ |LR\rangle -|RL \rangle \right]\equiv \Psi_{\mathrm{space}}^{(-)}$ &
$\dfrac{1}{\sqrt{2}}(\, \ket{\phi \tilde{\phi}} - \ket{\tilde{\phi} \phi} \,)$ \\
c & singlet & $+$ & $+$ & $\dfrac{1}{\sqrt{2}}\left[ \ket{LL} + \ket{RR} \right]$ & $\ket{\tilde{\phi} \tilde{\phi}}$ \\
d & singlet & $+$ & $-$ & $\dfrac{1}{\sqrt{2}}\left[ \ket{LL} - \ket{RR} \right]$ &
$\dfrac{1}{\sqrt{2}}(\, \ket{\phi \tilde{\phi}} + \ket{\tilde{\phi}
\phi}\,)$ \\
\hline
\end{tabular}
\caption{The four lowest energy eigenstates of the
two-atom double well system, as shown in Fig.~\ref{adiabatic_evolution}. Only states $a$ and $b$
are used in the quantum logic gate; these are degenerate in the
limit of an infinite barrier. (In this limit states $c$ and $d$ are also
degenerate, both having energy $U$ greater than $a$ and $b$.) As the
barrier height is decreased, $a$ and $b$ evolve into states with an
energy splitting between them and a phase difference accumulates
which is an essential part of the operation gate described in the
text. The localized eigenstates $\langle x | L\rangle$ and $\langle
x | R\rangle$ are not appropriate for a low barrier (high tunneling
rate between wells), and we should use products of the even and odd
parity wave functions $\phi$ and $\tilde\phi$ [see Eq.~(\ref{xLR})]. These states are included for completeness in the right
column. \label{abcdstates}}
\end{table}

\clearpage\section*{Figure captions}

\end{document}